# Molecular dynamics simulation for cross-linking processes and material properties of epoxy resins with the first principle calculation combined with global reaction route mapping algorithms

Yutaka Oya[a,*], Masahiro Nakazawa[a], Keiichi Shirasu[a], Yuki Hino[a], Kyosuke Inuyama[a], Gota Kikugawa[b], Jing Li[a], Riichi Kuwahara[c], Naoki Kishimoto[d], Hiroki Waizumi[d], Masaaki Nishikawa[e], Anthony Waas[f], Nobuyuki Odagiri[g], Andrew Koyanagi[g], Marco Salviato[h], Tomonaga Okabe[a,i]

[a]*Department of Aerospace Engineering, Tohuku University, 6-6-01 Aramaki-Aza-Aoba, Aoba-ku Sendai, Miyagi 980-8579, Japan*

[b]*Institute of Fluid Science, Tohuku University, 2-1-1 Katahira, Aoba-ku, Sendai, Miyagi 980-8577, Japan*

[c]*Dassault Systèmes K.K., ThinkPark Tower, 2-1-1 Osaki, Shinagawa-ku, Tokyo 141-6020, Japan*

[d]*Department of Chemistry, Tohuku University, 6-3 Aramaki Aza-Aoba, Aoba-ku, Sendai 980-8578, Japan*

[e]*Department of Mechanical Engineering and Science, Kyoto University, Nishikyo-ku 615-8540, Japan*

[f]*Composite Structures Laboratory, Aerospace Engineering, University of Michigan, Ann Arbor, MI 48109-2140, USA*

[g]*Toray Composite Materials America, Inc Tacoma, WA 98446, USA*

[h]William E. Boeing Department of Aeronautics and Astronautics, University of Washington, 98195-2400, WA, USA

[i]Department of Material Science and Engineering, University of Washington, USA

## Abstract

Herein, epoxy resin is cured by coupling quantum chemical (QC) calculations with molecular dynamics (MD) simulations that enable parameter-free prediction of material characteristics. A polymer network is formed by the reaction between base resin and curing agent. The reaction uses activation energy and heat of formation data obtained by first-principle calculations coupled with global reaction route mapping (GRRM) algorithms. Density, glass transition temperature, Young's modulus, and curing conversion is used to validate the procedure. Experimental and simulation results indicate that base resin with multi-functional reaction groups increases glass-transition temperature and Young's modulus because of cross-linked formations at the molecular scale.

**Keywords:** Epoxy resin, GRRM, Molecular Dynamics, QC, Simulation, Polymer

I. INTRODUCTION

Carbon fiber-reinforced plastic (CFRP) exhibits superior strength and stiffness while maintaining its light weight. In recent years, CFRPs have been applied to industrial fields such as aerospace engineering, civil engineering, and wind energy production [1-3]. Their application to structural components of aircrafts potentially reduces specific fuel consumption and increases the payload capacity [4]. Research and development of composite materials have been actively performed to improve their properties. Specifically, the thermosetting resin has been widely studied, which is used as a matrix material and controls the properties of composites.

Epoxy resin, which is the matrix generally used in CFRPs, can be produced using the chemical reaction between a base epoxy resin and a curing agent to form a three-dimensional cross-linked structure. In general, the appropriate selection of the thermosetting resin appears challenging because of the numerous combinations of curing agents and base epoxy resins.

Molecular Dynamics (MD) simulations have increased drastically in recent years and are presumably the ideal support tools to accelerate selections of the appropriate combination. Several studies for the modeling of thermosetting resin have been reported in the past decade. For example, Fan et al. simulated a curing process of epoxy resin using 628 atoms and then measured the glass transition temperature ($T$g) and the elastic modulus [5]. Similarly, Wu *et al*. predicted the $T$g and analyzed the temperature-dependent radial distribution functions and internal structures of epoxy resin [6]. Komarov et al. simulated a curing process of epoxy resin utilizing both the coarse-grained and all-atom models [7]. Varshney *et al*. conducted a non-equilibrium MD simulation to investigate the thermal conductivity of epoxy resin [8]. Li and Strachan reproduced the cured structure of epoxy resin using 15,198 atoms, predicted the $T$g and elastic modulus [9], and also investigated the relationship between strain rate and yield stress [10]. Furthermore, Li et al. predicted variations in the density ($\rho$), $T$g, and elastic modulus for epoxy resin with different compositions and their results agreed with experiment values [11]. Ghosh et al. performed MD simulation to obtain the phonon dispersion relation in epoxy and confirmed the influence of strain and stress gradients and their contribution near geometric discontinuities [12].

However, the previously mentioned studies did not appropriately consider the curing reaction of the base epoxy resin and the curing agent since only the distance between functional groups was used for the reactions. Recently, Estridge simulated the curing process considering ad-hoc reaction probabilities and pointed out that reaction probability affects the final cross-link structures [13]. Quantum Chemical (QC) calculation is generally required to simulate the reaction process. However, since the number of atoms in QC calculation is more limited than that in MD simulation, quantum MD (QMD) simulation is considered to be impractical for the polymer system. Consequently, the MD simulation that incorporates reaction energies obtained from quantum calculations was recently proposed to overcome such challenges [14, 15]. Okabe et al. [14] predicted curing characteristics for some combinations of base epoxy resins and curing agents; this was achieved by using the activation energy and the heat of formation obtained via the semi-empirical Molecular Orbital (MO) method. Moreover, Okabe et al. predicted the Young's modulus and ρ for the several epoxy resin systems [16,17]. Their predicted results were in accordance with the experimental results. Although they successfully predicted the properties in their study, the semi-empirical MO used in the literature is perceived as inaccurate for calculating the transition states (TSs). Hence, the more accurate first principle calculation is required.

 The purpose of this study is to refine our proposed simulation procedure by introducing the first principle calculation, which is combined with the global reaction route mapping (GRRM) algorithms [18,19] that enable the automatic exploration of TSs and energy the

associated with a chemical reaction of epoxy resin. The present paper is organized as follows:

Section II. Describes the details of the simulation method and the experimental procedures.

Section III. Presents the results and discusses the validation of the proposed approach as well as the process of the polymer network formation

Section III. Presents the conclusions obtained from this study.

## II. SIMULATION PROCEDURE AND EXPERIMENTAL METHODS

In this study, we investigated bisphenol A diglycidyl ether (DGEBA) and tetraglycidyl diaminodiphenyl methane (TGDDM) as epoxy resins and 4,4´-Diamino diphenyl sulfone (4,4´-DDS) as curing agents (Fig. 1). We systematically varied the molar ratio between DGEBA and TGDDM, as shown in Table 1. The bulk density, $T$g, Young's modulus and final curing conversion were estimated using the simulations and experiments.

Our simulation consists of three steps. In step one, the first principle calculation is conducted to evaluate the activation energies and the heat of formation for the MD simulation. The first principle calculation (Hartree-Fock method) is used along with GRRM algorithms to determine transition states. In step two, the previously calculated energies are used in the curing simulation. The proposed simulation can reproduce a polymer network, which cannot be achieved by independently using the QMD simulation; this is due to limitations of the computational cost. In step three, MD calculations are solved to obtain thermal and mechanical properties by using the obtained polymer networks. Details of the calculations are discussed in the following subsections. The experimental procedure used to obtain $T$g, Young's modulus, and curing conversion are also described at the end of this section.

### A. First Principle Calculation to Obtain Activation Energy and the Heat of Formation

First principle calculations were utilized to evaluate the activation energy and heat of formation. The calculations were performed using a quantum chemistry package program (Gaussian09 [20]) at the restricted Hartree-Fock SCF level of theory with a 3-21G basis set for all atoms (RHF/3-21G). The zero-point vibrational energies at the stationary points were also calculated. The reaction pathways from the epoxy reactants to the target products on the potential energy surface were searched using the GRRM program [19]. First, we used the sphere contraction walk (SCW) method [21] in the GRRM program to obtain intermediates on the reaction pathway. Then, we applied a two-point scaled hypersphere search (2PSHS) method [22] to two equilibrium structures (EQs) located just before and after the TS for the ring-opening reaction of the epoxy group.

As illustrated in Fig. 2, the primary amines ($-NH_2$) in the curing agent successively react with the epoxy group in the base resin. Thus, the primary amine becomes a secondary amine once it reacts with the epoxy group. Likewise, the secondary amine (-NH) changes to a tertiary amine, and no longer reacts. The calculated results for activation energy and heat of formation for the first and second reactions are listed in Table 2.

### B. MD Calculation for Cross-Linked Formation and Physical Properties

MD simulations were performed to simulate the curing process and analyzed physical properties. The number of atoms used in the simulation ranged from 9000 to 12700. We assigned the COMPASS II force field as the interaction potential for polymer materials, and utilized Materials Studio 8.0 to obtain the bulk density [23]. The TEAM force field [24] and the LAMMPS software package [25] were used to evaluate $T$g and Young's modulus.

To reproduce the curing reaction, we employed the Monte Carlo method [14, 15]. The criterion for generating cross-linking is determined by a reaction probability (k) based on

the Arrhenius equation:

$$k = A\exp\left(-\frac{E_a}{RT}\right), \tag{1}$$

where $E_a$ is the activation energy, $R$ is the gas constant, and $T$ is the local temperature around the reactive sites. In addition, $A$ can be defined as the acceleration constant, which is set to $10^{15}$ so that the reaction can be reproduced within a realistic computational time. The MC method is applied to all the functional groups within the distance of 5.64 Å [9]. After generating covalent bonds between the functional groups, the heat of formation, $\Delta H$, is assigned to the local kinetic energy as

$$K_{\text{after}}^{\text{mol}} = K_{\text{before}}^{\text{mol}} + \Delta H, \tag{2}$$

where $K^{\text{mol}}$ is the kinetic energy of the molecule, and the subscripts denote before and after the reaction. We performed the curing simulation at isothermal (453 K) and isopressure (1 atm) conditions and defined the curing conversion $\alpha$ by

$$\alpha \equiv \frac{f_{\text{reacted}}}{f_{\text{total}}} \times 100, \tag{3}$$

where $f_{\text{total}}$ is the total number of epoxy groups and $f_{\text{reacted}}$ is that of the reacted epoxy groups.

Physical properties were calculated for all samples at the final curing conversion. First, relaxation calculation was performed to obtain equilibrium states and their densities. Second,

$T_g$ and Young's modulus were evaluated. The calculation of density ρ requires structural optimization and dynamic calculation under the NPT ensemble. In general, the calculated density fluctuates due to pressure control. Therefore, average densities in the equilibrium state under conditions of 1 atm and 298 K were utilized for comparisons. We obtained $T_g$ from the relationship between volume and temperature at the NPT ensemble with the target temperature from 300 K to 600 K at a heating rate of 10 K/ns. Volume change was approximated by two straight lines using the least-square approximation, and the intersection point of those lines was defined as $T_g$.

Young's modulus is given as the slope of the stress-strain curves. The simulation cell was then elongated in the uniaxial direction using the NPT ensemble to obtain stress vs strain curves. The strain rate was set to $5.0 \times 10^8$ (1/s), and the simulation cell was deformed until the strain reached 3%. Since stress fluctuates in MD simulation, the obtained stress-strain curves were averaged with 15 different simulations.

### C.  Experimental Procedure

The cured samples were experimentally prepared by the following methods. In the case of DGEBA/4´4-DDS (sample 1), a base resin and a curing agent were mixed by stirring for 6 minutes and degassing for 30 minutes using a mixer (Tninky Corp). This sample was then heated at 180 °C and 1 atm for 8 hours to generate the cured sample. However, the stirring and degassing processes in other samples were different from those in sample 1 because these samples contain TGDDM of high viscosity. First, TGDDM was heated to decrease its viscosity, thereby facilitating mixing with DGEBA and 4´4-DDS. The mixture was stirred for 10 minutes and degassed for 2 minutes and then degassed at 100 °C and 0.008 MPa for 1 h before heating.

The density of the cured sample was measured based on Archimedes' principle by $\rho = \frac{A(\rho_0 - \rho_L)}{A-B} + \rho_L$, where $A$ is the weight of a sample under atmosphere, $B$ is the weight of a sample in distilled water, $\rho_0$ is the density of distilled water at room temperature, and $\rho_L$ is atmospheric density. $A$ is measured after sufficient drying at 100 °C using a balance (Balance XS 104 by Mettler-Toledo, Inc.)

$T$g of a cured sample was measured by dynamic mechanical analysis (DMA 7100 by HITACHI High-Technologies. Inc.). To obtain $T$g, the temperature was changed while applying stress at a constant frequency to the sample. The temperature at the peak loss tangent is defined as $T$g. In this study, the temperature of the sample was increased from 30–300 °C at a heating rate of 2 °C/min while applying the stress at 1 Hz to the sample.

A stress-strain curve was measured by a uniaxial tensile test using a tensile testing machine (MTS Landmark 810). The tensile rate is 1 mm/min, in accordance with the standard based on JIS K 7161-2. Strain gauges (Kyowa Electronic Instruments Co., Ltd.) were attached to the dog-bone shaped specimen, following the IBA standard. A tensile test was conducted 5 times for each resin, and the averaged values were used for comparison with experiments.

The curing conversion of the cured sample was evaluated by Fourier transform infrared spectroscopy (FT-IR, Nicolet iS20 by Thermo Fisher SCIENTIFIC). The amount of functional groups is determined from the intensity of the infrared absorption peak. The curing conversion is calculated by comparing the spectrum intensities before and after curing. However, the spectrum varies significantly depending on the shape of the sample. Thus, the spectrum intensity of the epoxy group is normalized with the intensity of a benzene ring that remains constant before and after curing. We attempted two criteria to evaluate spectrum intensity, *i.e.*, height and area of spectrum intensity. The spectrum peak for the epoxy group is found between 870 cm$^{-1}$ and 930 cm$^{-1}$, and the peak for the benzene ring is found between 770 cm$^{-1}$ and 870 cm$^{-1}$. As shown in Fig. 3, we made baselines by connecting the local minimums at both ends of a peak. We determined the height of the intensity ($h$) as the height from the baseline to the peak. The area of the intensity ($a$) is determined by the area of the region surrounded by the baseline and the spectrum. The number ratio of the epoxy group and benzene ring $r_{\text{total}}$, and the number ratio of the reacted epoxy group and the benzene ring $r_{\text{reacted}}$ were obtained by $r_{\text{total}} = h_{\text{epoxy}}^{\text{before}}/h_{\text{benzene}}^{\text{before}} = a_{\text{epoxy}}^{\text{before}}/a_{\text{benzene}}^{\text{before}}$ and $r_{\text{reacted}} = \left(1 - h_{\text{epoxy}}^{\text{after}}\right)/h_{\text{benzene}}^{\text{after}} = \left(1 - a_{\text{epoxy}}^{\text{after}}\right)/a_{\text{benzene}}^{\text{after}}$, where subscripts "before" and "after" represent before and after curing reaction, respectively. The curing conversion $\alpha$ was calculated by $\alpha = r_{\text{reacted}}/r_{\text{total}}$.

## III. RESULTS AND DISCUSSION

To validate the simulations, we compared the density, $T$g, Young's modulus, and curing conversion obtained from simulation with those of experiment. The obtained experimental data for density, $T$g and Young's modulus of DGEBA/4,4´-DDS (sample 1) and TGDDM/4,4´-DDS (sample 7) are consistent with previous experiments [14, 25]. Fig. 4 shows that $T$g and Young's modulus linearly increase with the amount of TGDDM for both the simulation and the experiment. Our simulated density does not change for the amount of TGDDM, although density obtained from the experiment slightly increases. Our simulation quantitatively reproduces both $T$g and Young's modulus. Further, the difference between simulated density and the experimental density is within 15%. Fig. 5 represents the comparison of final curing conversions obtained by our simulation and the experiment. All samples achieve ~80% conversions. Therefore, the simulated results agree with the experimental results. These comparisons indicate that our simulation can reproduce realistic polymer network structures.

Molecular scale branching structures are used to explain why TGDDM improves $T$g and Young's Modulus (Fig. 6). In Fig. 6(a), the circles denote the sites that have not yet reacted, whereas the line denotes reactive functional group with another type of functional group. Fig. 6(b) depicts the typical network in TGDDM/4,4´-DDS.

Fig. 7(a) and (b) illustrate plots of the distribution of branching structures versus the curing conversion for DGEBA/4,4´-DDS (sample 1) and TGDDM/4,4´-DDS (sample 7). The ratio is calculated by the number of monomers with branching structures as a fraction of the total number of monomers. Two-branching structures continue to increase and remain dominant among the possible branching structures until the end of the reaction when DGEBA is used as the base resin. For TGDDM/4,4´-DDS, three-branching and four-branching structures become dominant after 60% conversion. On the other hand, the zero-branching structure disappears and the one-branching and two-branching structures decrease, which suggests that the reaction type changes from intermolecular to intramolecular reactions around 60% conversion. Fig. 7(c) shows the distribution of branching structures with respect to the amount of TGDDM. One-branching structure and two-branching structure linearly decrease with the amount of TGDDM, while the three-branching structure and four-branching structure increase. These results suggest that TGDDM produces densely cross-linked structures through an intramolecular reaction at the latter stage of curing reactions. The densely cross-linked structure also makes glass-transition temperature and Young's modulus increase as shown in Fig. 4.

## IV. CONCLUSION

We proposed a simulation procedure for the curing reaction of epoxy resin by coupling first principle QC calculations with MD simulations. Specifically, we refined our previous approach by introducing the first principle method combined with GRRM algorithms. Our simulated results are quantitatively validated by experiments that determine the curing conversion, density, Young's modulus, and $T$g for DGEBA/TGDDM/4,4´-DDS mixtures. Our simulation indicates that the number of reactive functional groups in a monomer greatly changes its cross-linked formation. The base resin with four functional groups (TGDDM) undergoes a curing process that changes from an intermolecular to an intramolecular reaction. However, for base resins with two functional groups (DGEBA), the intermolecular reaction occurs throughout the curing process. We conclude that thermal and mechanical properties are affected by branching structures. Thus, the intramolecular reaction enhances the rigidity of the resin network, so that Young's modulus and the glass transition temperature of TGDDM with high dense cross-links or branches surpass those of DGEBA. In summary, the simulation procedure proposed in this study is a powerful tool for parameter-free prediction of the curing process and thermal and mechanical properties.

## ACKNOWLEDGMENTS

This work was supported by the Council for Science, Technology and Innovation (CSTI), the Cross-ministerial Strategic Innovation Promotion Program (SIP), and "Materials Integration" for the revolutionary design system of structural materials (Funding agency: JST). We also acknowledge the financial support of Toray Composite Materials America, Inc. N.K. and T.O. acknowledge a research grant from the Institute for Quantum Chemical Exploration (IQCE). The authors would like to acknowledge the vitally important encouragement and support provided by the University of Washington-

Tohoku University: Academic Open Space (UW-TU: AOS).

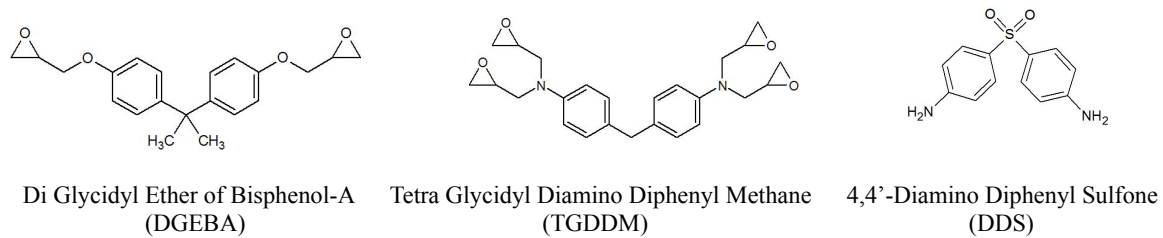

Di Glycidyl Ether of Bisphenol-A    Tetra Glycidyl Diamino Diphenyl Methane    4,4'-Diamino Diphenyl Sulfone
(DGEBA)                            (TGDDM)                             (DDS)

Fig. 1. Molecular structures of base resins and curing agents that constitute target materials.

Table 1. Samples and molar ratios used for the MD simulation.

| Sample No. | Molar ratio [DGEBA / TGDDM / 44DDS] | Ratio of TGDDM against base resins [%] |
|---|---|---|
| 1 | 2 / 0 / 1 | 0 |
| 2 | 6 / 1 / 4 | 25 |
| 3 | 4 / 1 / 3 | 33 |
| 4 | 2 / 1 / 2 | 50 |
| 5 | 2 / 2 / 3 | 67 |
| 6 | 2 / 3 / 4 | 75 |
| 7 | 0 / 1 / 1 | 100 |

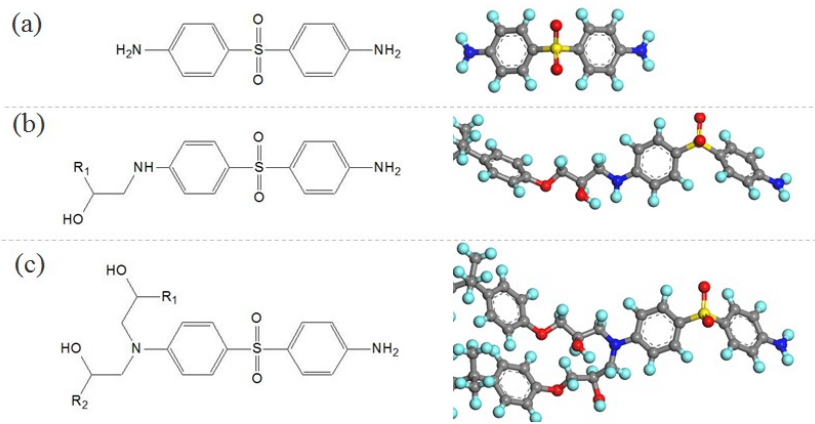

Fig. 2. Classification of reactions in amino groups of the curing agent. (a) No reaction. (b) First reaction. (c) Second reaction. The chemical structure is on the left, and the atomistic models for 4,4´-DDS are on the right.

Table 2. Calculated energies for the primary and secondary curing reactions using the first principle calculation

| 1st reaction | 2nd reaction | $E_a$ [kcal/mol] | $H_f$ [kcal/mol] |
|---|---|---|---|
| DGEBA | - | 45.59 | 40.98 |
| DGEBA | DGEBA | 43.04 | 49.96 |
| DGEBA | TGDDM | 50.18 | 41.50 |
| TGDDM | - | 38.42 | 48.08 |
| TGDDM | TGDDM | 50.54 | 28.53 |
| TGDDM | DGEBA | 47.87 | 27.66 |

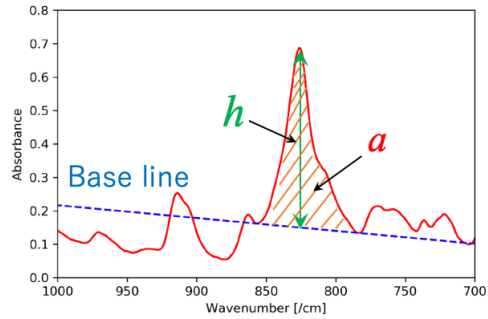

Fig.3 Baseline for the height and the area of spectrum intensity.

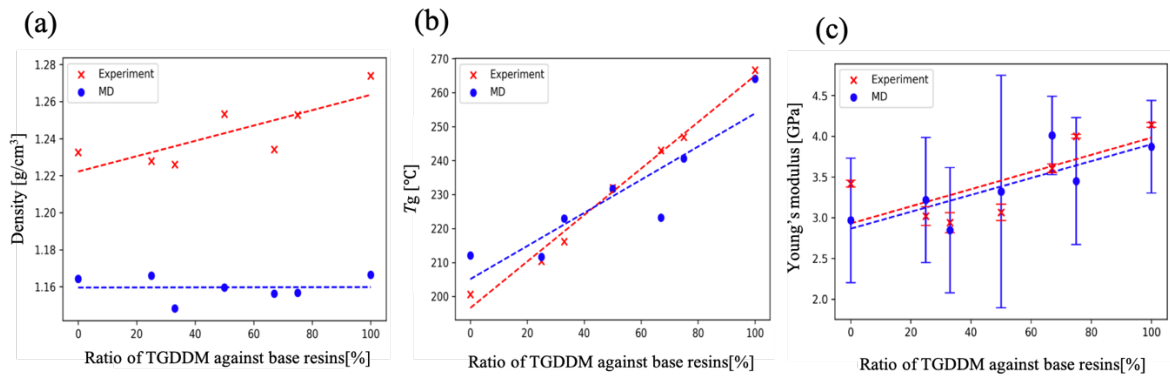

Fig. 4. Material characteristics depending on the molar rate of TGDDM. (a) Density. (b) Glass transition temperature. (c) Young's modulus.  MD results (● blue) and experimental results (× red). Dashed lines are approximated straight lines obtained by the least square method.

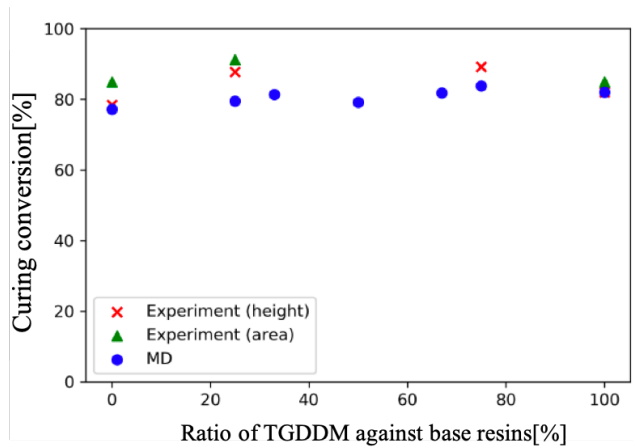

Fig. 5. Final conversion rate with respect to the molar rate of TGDDM. MD results (● blue), FT-IR results with height criteria (× red) and FT-IR results with area criteria (▲ blue curve) are presented.

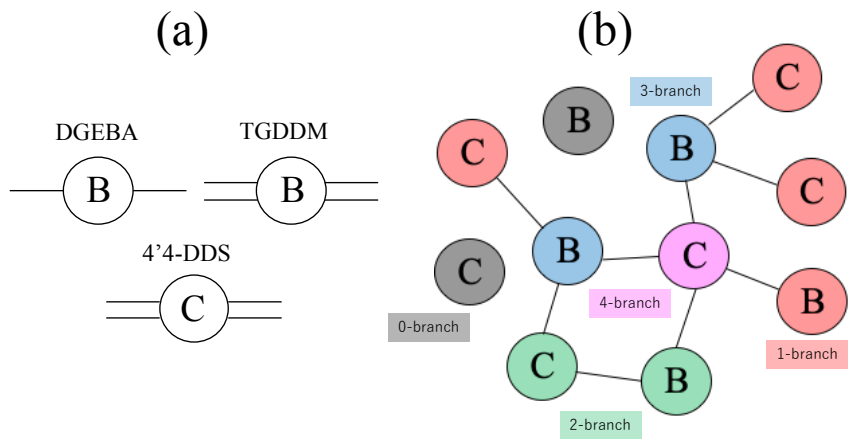

Fig. 6. (a) Symbols of base resins and curing agents. The number of lines represents the maximum number of reactions. (b) Network structure formed by base resin and curing agent. Colors correspond to the number of branches: black for zero-branching structure, red for one-branching structure, green for two-branching structure, blue for three-branching structure, and purple for four-branching structure.

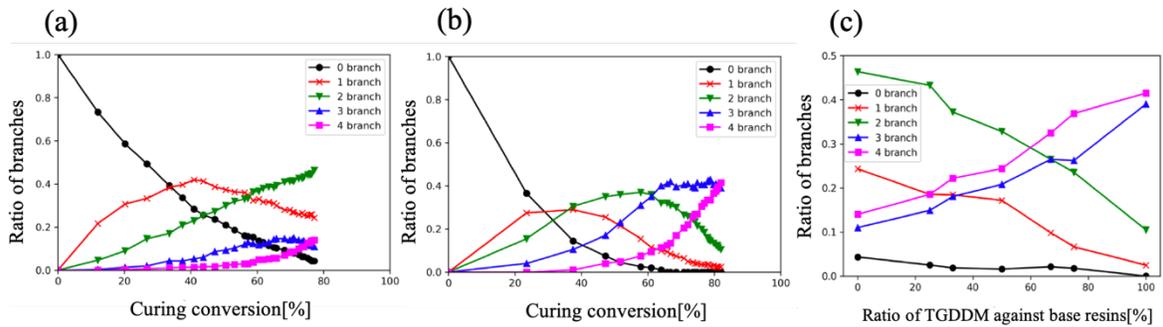

Fig. 7. Branch ratios depending on curing conversion for (a) DGEBA/44-DDS (sample 1). (b) TGDDM/4'4-DDS (sample 7), and (c) brach ratio for all samples at final conversion. Zero-branched (● black curve), one-branched (× red curve), two-branched (▼ green curve), three-branched (▲ blue curve), and four-branched (■ purple curve) structures are presented. Colors correspond to those in Fig. 6 (b).